\documentclass[sigconf]{acmart}
\usepackage{amsmath,amssymb,amsfonts}
\usepackage{algorithmic}
\usepackage{textcomp}
\usepackage{xcolor}
\usepackage{soul}
\usepackage{url}
\usepackage{tabularx}
\usepackage{subfig}
\usepackage{textcomp}
\usepackage{multirow}
\AtBeginDocument{%
  \providecommand\BibTeX{{%
    \normalfont B\kern-0.5em{\scshape i\kern-0.25em b}\kern-0.8em\TeX}}}


\copyrightyear{2020}
\acmYear{2020}
\setcopyright{acmlicensed}\acmConference[SIGIR '20]{Proceedings of the 43rd International ACM SIGIR Conference on Research and Development in Information Retrieval}{July 25--30, 2020}{Virtual Event, China}
\acmBooktitle{Proceedings of the 43rd International ACM SIGIR Conference on Research and Development in Information Retrieval (SIGIR '20), July 25--30, 2020, Virtual Event, China}
\acmPrice{15.00}
\acmDOI{10.1145/3397271.3401066}
\acmISBN{978-1-4503-8016-4/20/07}
\begin{document}
\fancyhead{}

\title{Modeling Personalized Item Frequency Information for Next-basket  Recommendation}


% \author{Haoji Hu$^1$, Xiangnan He$^2$*, Jinyang Gao$^3$, Zhi-Li Zhang$^1$}
% \affiliation{
%   \institution{$^1$University of Minnesota Twin Cities, $^2$University of Science and Technology of China, $^3$Alibaba Group}
% }
% \email{huxxx899@umn.edu,  xiangnanhe@gmail.com,  jinyang.gjy@alibaba-inc.com, zhzhang@cs.umn.edu}

\author{Haoji Hu}
\affiliation{%
  \institution{University of Minnesota, Twin Cities}
}
\email{huxxx899@umn.edu}

\author{Xiangnan He}
\authornote{Xiangnan He is the corresponding author.}
\affiliation{%
  \institution{University of Science and Technology
of China}
}
\email{xiangnanhe@gmail.com
}

\author{Jinyang Gao}
\affiliation{%
  \institution{Alibaba Group}
}
\email{jinyang.gjy@alibaba-inc.com}

\author{Zhi-Li Zhang}
\affiliation{%
  \institution{University of Minnesota, Twin Cities}
}
\email{zhzhang@cs.umn.edu}

\renewcommand{\shortauthors}{.}

\begin{abstract}
Next-basket recommendation (NBR) is prevalent in e-commerce and retail industry. In this scenario, a user purchases a set of items (a basket) at a time. NBR performs sequential modeling and recommendation based on a sequence of baskets. NBR is in general more complex than the widely studied sequential  (session-based) recommendation which recommends the next item based on a sequence of items. Recurrent neural network (RNN) has proved to be very effective for sequential modeling, and thus been adapted for NBR. However, we argue that existing RNNs cannot directly capture item frequency information in the recommendation scenario. 

 Through careful analysis of real-world datasets, we find that {\em personalized item frequency} (PIF) information (which records the number of times that each item is purchased by a user) provides two critical signals for NBR. But, this has been largely ignored by existing methods. Even though existing methods such as  RNN based methods have strong representation ability, our empirical results show that they fail  to learn and  capture PIF. As a result, existing methods cannot fully exploit the critical signals  contained in PIF. 
 Given this inherent  limitation of RNNs,  we propose a simple item  frequency based  k-nearest neighbors (kNN) method to directly utilize these critical signals. 
 We evaluate our method on four public real-world datasets. Despite its relative simplicity, our method frequently outperforms the state-of-the-art NBR methods -- including deep learning based methods using RNNs -- when patterns associated with PIF play an important role in the data. 
\end{abstract}





\maketitle

\section{Introduction}

Recommendation systems have  been applied in many different  applications~\cite{aggarwal2016recommender}. NBR is a  type of  recommendation  problem that aims to recommend a set  of items to a user based on his/her  historical  purchased  baskets~\cite{rendle2010factorizing}\cite{ying2018sequential}\cite{wang2015learning}\cite{yu2016dynamic}, which is prevalent in E-commerce and retail industry.  Unlike top-$n$  recommendation (whose historical  record is a set of items)~\cite{ning2015comprehensive} and sequential recommendation (whose historical record is a sequence of items)~\cite{ludewig2019performance}, the historical record of the next-basket  recommendation is a sequence of sets or sequential sets (whose element is a set). Considering the historical  records, top-$n$  recommendation and sequential/session-based recommendation can be seen as special cases of NBR when the NBR only has one  basket and has a sequence  of baskets whose size are all of 1, respectively. But in recommendation step, top-$n$ recommendation only recommends  new items that are not  contained in the user's historical records,  whereas both  sequential/session-based  recommendation and NBR  recommend new and old items. Even though sequential/session-based recommendation is similar to NBR, we cannot directly  apply  sequential/session-based  recommendation method to do  NBR without messing up the  information  existing in the  sequential  sets\footnote{For example, we can convert each basket into a sequence and concatenate the sequences from different baskets in temporal order. This  introduces a non-existing order among items within the same basket.}.

The challenging part in NBR is how to model the relation between the  historical records and recommended  items. Existing NBR methods use  different ways to  model the information in the  historical records as the  user profile and capture user-item  interactions for predicting the next basket. RNNs have become one of the mainstream choices as it is easy to be  tweaked for sequential modeling.  However, we argue that existing RNNs  cannot directly capture item frequency  information in the recommendation  scenario. 

Recently, repeated behaviors are found to bring  considerable  performance  improvement in both sequential/session-based recommendation~\cite{wang2019modeling}\cite{ren2019repeatnet} and NBR~\cite{hu2019sets2sets}. It is based on the observation that repeated purchases widely exist in the users' records. However, our analysis shows  that PIF contains more information than repeated  purchase pattern. We  observe that similar users' PIF also contains  collaborative purchase pattern. This new pattern shows that if a user repeatedly purchases a  item, similar users are likely to purchase the same  item. Existing methods fail to fully utilize this  useful information  contained in PIF. 

In this paper, 
we propose a  simple k-nearest  neighbors (kNN)  based method which directly captures the two useful patterns associated with PIF. To demonstrate the  effectiveness, we  also observe and  analyze the  limitation of  existing  methods to capture the  important patterns associated with PIF as they cannot learn the vector  addition well. In  summary,  our  contributions are as follows:
\begin{itemize}
    \item We analyze two  patterns associated with PIF and the target basket. The collaborative purchase pattern that PIF can contribute to the NBR  in a  collaborative way is discovered. 
    \item We discover the  difficulty of RNNs in learning  vector addition in  recommendation setting.  To our best  knowledge, we  are the first to present and  analyze this  phenomenon. 
    \item We propose a  simple and effective kNN based method to directly capture the two useful patterns  associated with PIF.  The  temporal  dynamics is also considered in the proposed method.  
    \item We  perform  experiments on four real-world data sets to  demonstrate the  effectiveness  of the proposed method.    
\end{itemize}

The rest of the paper is organized as follows: 
Section~\ref{sec:preliminaries} presents the preliminaries.  In section~\ref{sec:method}, we describe our proposed method.
In section~\ref{sec:related_work}, we discuss the related work. In section~\ref{sec:experiment}, we evaluate our method. Section~\ref{sec:conclusion} provides some concluding remarks and future directions.

\section{PRELIMINARIES}
\label{sec:preliminaries}
In this section, the NBR  problem is first formally defined. Next, we analyze the  patterns associated with PIF for NBR. Our analysis towards  real-world data sets  reveals that PIF contains  two important signals for  the next target basket.  Finally, we summarize the  existing methods, and  discuss the their difficulty in learning PIF.   

\subsection{Problem Definition} 

Given the historical  purchase records of a user $\small\{\mathbf{v}_{1}, \mathbf{v}_{2}, ..., \mathbf{v}_{i}, ...,  \mathbf{v}_{t}\}$, where a  set of items (a basket) at the $i$-th time step is represented as a 0/1 vector $\small\mathbf{v}_{i}$ whose entry $c_j$ ($j\in[0,d]$) is set to 1 if the  corresponding item  appears in the  basket, our  goal is to predict the next set of items (next basket)  $\small\mathbf{\hat{v}_{t+1}}$. 

Following the  literature~\cite{rendle2010factorizing}\cite{yu2016dynamic}, we consider a  fixed-size set with $s$ items as the  recommendation for the next basket.

% \subsection{Data Sets}
% Existing methods for top-$n$ recommendation and sequential/session-based  recommendation  are usually evaluated on several public data sets (e.g. MovieLens 1M\footnote{https://grouplens.org/datasets/movielens/1m/}and Netflix~\cite{bennett2007netflix}). We observe that existing methods are evaluated on different data sets. For example, the classical NBR method FPMC~\cite{rendle2010factorizing} is evaluated on two private data sets.  

\subsection{Relation between PIF and Target Basket}
\label{sec:analysis_freq}

In this section, we discuss two important patterns  associated with PIF that can help predict the  target next basket: \textbf{repeated purchase pattern} and \textbf{collaborative purchase pattern}. 

Repeated purchase behavior
in grocery shopping and online activities has been studied in the areas of economics and psychology theoretically and empirically~\cite{anderson2014dynamics}\cite{benson2016modeling}\cite{dawes2015has}\cite{jacoby1973brand}.  This pattern has been used in recent sequential/session-based recommendation method~\cite{ren2019repeatnet}\cite{wang2019modeling} and NBR method~\cite{hu2019sets2sets}\cite{wan2018representing}, and is shown to get  considerable performance gain.  Specifically, a simple baseline which recommends the  user-specific most  frequent items can sometimes outperform many existing  methods~\cite{hu2019sets2sets}.  The good performance of this baseline comes from the assumption that the next  target basket often  contains items that have appeared in the user's historical records. And the  higher PIF is associated with a higher probability of the corresponding item  to  appear in the target basket again.  However, this assumption has a limitation that the PIF only helps the target  user. It ignores the collaboration among different users. This is  the core idea of  collaborative filter~\cite{konstan1997grouplens}. A natural question is whether PIF can help in  a collaborative manner. To  verify this, we investigate the co-occurrence of the same item to simultaneously appear in the past baskets of the similar users and  the next basket of the  target user on four  real-world data sets (the details of the data are in  section~\ref{sec:exp_data}).  To compare with repeated purchase pattern,  we also investigate the  co-occurrence of the same item to simultaneously  appear in the past baskets of the target user and the  next basket of the target user. We simply use the PIF vector ($\small PIF_{vector} = \sum_{i=1}^t\mathbf{v}_i$) to represent each user. For each user, his/her top 300 nearest neighbors (the total number of users in all data sets is no less than 10,000) are found based on the PIF vector. Denote the set of all items in target user's past baskets as $P$. Denote the set of all items in the neighbors' past baskets as set $N$. The target basket is denoted as $T$.  We calculate the  following four ratios: 
\begin{itemize}

    \item{$Recall_P$:} The  average recall of using $P$  to retrieve items in $T$.  Formally, $\small Recall_P=\frac{|P\cap T|}{|T|}$. It represents the  ratio of items captured by  repeated purchase pattern.
    \item{$Recall_N$:} The  average recall of using $N$  to retrieve items in $T$.  Formally, $\small Recall_N=\frac{|N\cap T|}{|T|}$. It represents the ratio of items captured by collaborative purchase pattern.
    \item{$Recall_{P\cap N}$:} The  average recall of using $P\cap N$  to retrieve items in $T$.  Formally, $\small Recall_N=\frac{|P\cap N\cap T|}{|T|}$. It represents the ratio of items captured by both repeated purchase pattern and collaborative purchase pattern.
    \item{$Recall_{\overline{P}\cap \overline{N}}$:} The  average recall of using $\overline{P}\cap \overline{N}$ to retrieve items in $T$.  Formally, $\small Recall_{\overline{P}\cap \overline{N}}=\frac{|\overline{P}\cap \overline{N}\cap T|}{|T|}$. It represents the  ratio of items not captured by  repeated purchase pattern and collaborative purchase pattern.

\end{itemize}
   
\begin{table}[ht]
\small
\centering
\caption{The importance of two patterns.}
\begin{tabular}{lcccc}
\toprule
\textbf{Data}&$Recall_P$ &$Recall_N$&$Recall_{P\cap N}$&$Recall_{\overline{P}\cap \overline{N}}$\\
\midrule

\emph{ValuedShopper}&0.6570& 0.9808&0.6490&0.0111 \\

\emph{Instacart}&0.5711 &0.8338& 0.5056&0.1007\\

\emph{Dunnhumby} &0.2777 &0.5580&0.2432&0.4075\\

\emph{TaFeng} &0.1378&0.8614 &0.1256 & 0.1262\\

\bottomrule
\end{tabular}
\label{tab:jaccard}
\end{table}

From Table  ~\ref{tab:jaccard}, we can make several observations. First, $Recall_P$ indicates that the repeated purchase pattern plays a considerable role in four data sets but varies dramatically across different data sets. Second, $Recall_N$ indicates that the collaborative purchase pattern plays much more  important role in all four data sets. It can help retrieve more than half items in the target basket. Surprisedly, this ratio can increase to more than 0.8 in three data sets. Note that, here we only use 300 nearest neighbors. It is expected that  $Recall_N$ will increase if we increase the number of neighbors. Third, $Recall_{P\cap N}$ indicates that these two patterns have overlap. Based on $Recall_N-Recall_{P\cap N}$, we can also infer that the collaborative purchase pattern provides extra signal related to the target basket. Fourth, $Recall_{\overline{P}\cap \overline{N}}$ indicates that only a small part of items are not covered by the two patterns. It implies that the  unseen patterns are  only a small part in the next basket.  Based on $1-Recall_{\overline{P}\cap \overline{N}}$, we find  that combining both patterns can achieve  better performance than any single pattern. Note that above analysis is based on complete $P$ and $N$ whose size is still large. In general, we only recommend a small number of items in NBR. Nonetheless this analysis demonstrates the  potential incorporating  PIF in NBR.

\subsection{Existing  NBR  Methods}
\label{sec:existing_NB}
\textbf{MC based methods:}
Rendle et al.~\cite{rendle2010factorizing} propose the classical NBR method which is based on factorization and Markov chain. Their method models the  personalized  item-item transition matrix between any pair of  consecutive baskets. Wang et al.~\cite{wang2015learning} propose a similar Markov chain model. Instead of using tensor  factorization, they  propose to use pooling to  aggregate the items  in the recent  basket as the recent  basket  representation and  predict the next  basket based on the aggregated  representation.  Ying et al.~\cite{ying2018sequential} enhance the structure  in~\cite{wang2015learning} and use   attention mechanism  to replace the  pooling operation.  The attention mechanism can focus on the most relevant items, which brings  performance  improvement. Also, they partition  historical items  into two sets. The items in the recent  basket represent  the short-term set and the items in the baskets before the recent one represent  the long-term set. Separated attention mechanisms are applied on both  sets to generate the hybrid user  representation. The prediction is based on the hybrid user representation and the item embedding. \\
\textbf{RNN based methods:} The assumption behind MC based methods is that the next basket is mainly decided by the last (or few last) baskets. However, MC based methods miss to capture the high-order dependency from long time ago. In order to capture the whole historical baskets,  RNNs are used to model the whole  history~\cite{yu2016dynamic}\cite{hu2019sets2sets}. Both of them use the same structure as ~\cite{wang2015learning} at each time step. The item embedding is first aggregated to generate the basket representation and then a RNN is used to model the temporal relation across all the  baskets. The hidden state of the last step of RNN is the user representation. And the next basket is predicted based on the generated user representation and target item embedding. Sets2Sets~\cite{hu2019sets2sets} also uses attention  mechanism to focus  on the most relevant baskets.

\subsection{Difficulty in Learning Item Frequency}

As PIF contains critical  information for NBR, an  immediate question is: can existing  methods capture this  information? We argue that whether  existing methods cannot capture this information, it will be hard for them to exploit this  critical information. Formally, we investigate if existing methods can learn the result of vector  addition  $\small\sum_{i=1}^t\mathbf{v}_i$ given the purchase recrods of a user. It is obvious that MC based methods cannot capture PIF, which is a type of high-order information, as MCs  only record last or last few  baskets. RNN based methods  have strong representation ability as RNNs can approximate any  computable  function~\cite{siegelmann1992computational}.  If RNNs can  aggregate the vectors in the same way that vector addition aggregates the  vectors, the last hidden state of the RNNs should contain the PIF information\footnote{Vector addition is a  more general case than two numbers addition which is shown in https://machinelearningmastery.com/learn-add-numbers-seq2seq-recurrent-neural-networks/. To our best knowledge, RNN is the most direct way for deep model to learn this operation as the number of vectors varies and the operation is repeated.}. In this section, we investigate if RNN can learn PIF. 

In the following, we  demonstrate that it is hard  for RNNs to learn PIF due to the  difficulty in the optimization. Our demonstration is as follows: First, we  analyze that the phenomenon is  related to the difficulty of searching the global  optimal solution for RNNs. As many  elements can lead to this problem, we  approach this  by eliminating other possible  causes. Second, we derive  a closed-form solution for  RNNs to learn the vector addition. Based on these two steps, we conclude that  even though RNNs have the general ability to learn, the training process sticks  into a local  minimum in current  recommendation setting. Finally, we discuss if we can overcome this  difficulty.

\subsubsection{Difficulty in Learning Vector Addition}

We use a  synthetic data set to illustrate this phenomenon. (1) We generate  2500 sequences of vectors  as the training data set. The dimensionality of all the vectors is  100. Each vector is a one-hot vector. The reason we only generate one-hot vectors is that any $q$-hot vector can be converted into $q$ one-hot vectors. But if we generate $q$-hot vectors, we cannot simulate the  case of  addition for  one-hot  vectors.  Thus, it is the simple but  general case.   Each sequence of vectors represents the vectors to be summed up. For simplicity, we fix the length of all the sequences  as 10. Fixing the length to  10 can also avoid the  difficulty  of RNNs in handling long-term dependency~\cite{pascanu2013difficulty}. (2) To make sure we obtain vectors that have  repeated items, we randomly select 2 out of the first 8 vectors as the last two vectors in each sequence of vectors. 

\begin{figure}[!t]
\centerline{\includegraphics[width=0.4\textwidth,height=0.2\textheight]{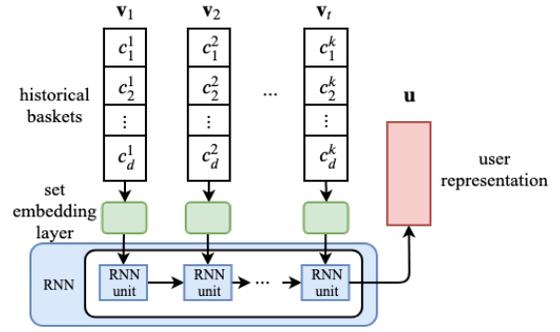}}
\caption{Existing RNN based NBR Module.}
\label{fig:RNN_framework}
\end{figure}

\begin{figure}[!t]
\centerline{\includegraphics[width=0.45\textwidth,height=0.13\textheight]{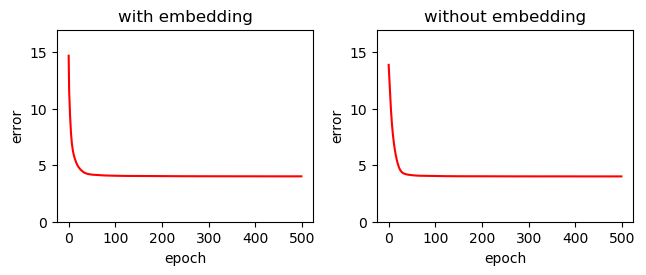}}
\caption{The training loss of using the component from Figure~\ref{fig:RNN_framework} to learn vector addition. The embedding size and RNN units are set to 64 (as our item space is 100, 64 is  the max value we tried for compressed representation setting to handle the sparsity in the input data).  Batch size is set to 64. The training loss is  the average mean square error (MSE) between the output of the last step (the final predicted sum vector) and the ground truth (the real sum vector).  GRU is used. Adam~\cite{kingma2014adam} is used for optimization. The learning rate is set to 0.001. All the parameters of different layers are randomly initialized by the default setting of Pytorch.}
\label{fig:adding_with_embedding}
\end{figure}

Existing RNN based method~\cite{hu2019sets2sets}\cite{yu2016dynamic} use a common  module to aggregate the historical baskets as a user representation in  Figure~\ref{fig:RNN_framework}.  Each basket is first input into the set embedding layer and then  transformed into set embedding. After that, a RNN is used to aggregate  all the set  embeddings at different time steps to  generate the final  user representation as a  summarized vector. This is the only part that has the potential to learn PIF as  it goes through all the past records and aggregates them into a user  representation. So we apply this module to learn vector addition on the synthetic data set. As only one-hot vectors are generated  in the data,  the set embedding  layer is reduced to an item embedding layer that is widely used existing deep learning based recommendation methods.  Note that the item embedding,  which is the input of the RNN,  and hidden state of RNN unit are usually of much smaller dimensionality  (usually is $2^z\in[8,  128], z\in\mathbb{Z}$) compared to the original one-hot vector (whose dimensionality is of  at least several thousands). This can help avoid the   parameter exploding and resolve the sparsity in the original one-hot vector space. So we  need to  project user  representation back to the original space to get the  predicted sum vector for  $\sum_{i=1}^t\mathbf{v}_i$. The training process is shown in  Figure~\ref{fig:adding_with_embedding}. The training loss converges to 4 which is far from the optimal error 0. To further show  this is a large  training error, we consider a very simple baseline that  directly predicts all the sum vectors as zero vectors. This baseline can  achieve an average MSE of  $\frac{(1-0)^2*8+(2-0)^2*2)}{10}=1.6$.  Usually, we may  speculate that the  embedding results in information loss. Thus, we remove the  embedding layer and directly forward the one-hot vector as the input of the RNN unit. Now the module shown in the  Figure~\ref{fig:RNN_framework} is simplified to a RNN. But this yields a similar  training error. Considering that optimizer may also affect the results,  we also check other two  widely-used  optimizers  SGD~\cite{bottou2010large} and RMSprop\footnote{\url{http://www.cs.toronto.edu/ tijmen/csc321/slides/lecture\_slides\_lec6.pdf}
}. However, the training  error does not change with different  optimizers. 

% \begin{figure}[!t]
% \centerline{\includegraphics[width=0.5\textwidth,height=0.14\textheight]{fig/adding_ten_vectors.png}}
% \caption{The training loss  with  different optimizers.  The learning rate is set to 0.001 for all the optimizers. Smaller learning rate is also explored but it only results in slower convergence.  Other  configurations are set to the  default values  provided by the toolbox in the Pytorch. Different  parameters for optimizers are explored, but the results are similar.}
% \label{fig:adding_ten_vectors}
% \end{figure}

A common concern for the failure of deep learning methods is that we do not  have enough data for training. By increasing  the training set size,  the training loss is expected to decrease. To verify this, we double the data size. We generate 2500 additional sequences of vectors and merge them with  the previous training data. However, the converged training error still remains the same. 

Another common speculation for large training error  is that the model's  capability is not enough. We should continue to increase the dimensionality.  However, we  argue small dimensionality is able to learn the optimal solution as we will  present a closed-form  solution that is not related to the  dimensionality in the next subsection.

% \begin{figure}[!t]
% \centerline{\includegraphics[width=0.18\textwidth,height=0.13\textheight]{fig/double_data.png}}
% \caption{The training loss after the training set is doubled.}
% \label{fig:doubled}
% \end{figure}

\subsubsection{Closed-form Solution for Vector Addition with RNN}

\begin{figure}[!t]
\centerline{\includegraphics[width=0.24\textwidth,height=0.13\textheight]{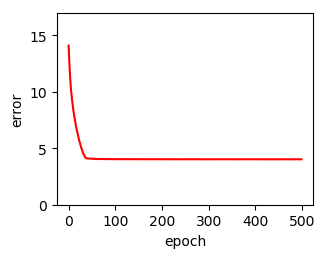}}
\caption{Without nonlinear activation functions.}
\label{fig:no_nonlinear}
\end{figure}

There are many different variants of RNNs~\cite{greff2016lstm}. Vanilla  RNN~\cite{mandic2001recurrent}, long short-term memory (LSTM)~\cite{hochreiter1997long} and  gated recurrent unit (GRU)~\cite{cho2014properties} are the most popular ones. LSTM and GRU are the extensions of vanilla RNN for handling long-term dependency  problem. For simplicity, we focus on the vanilla RNN as other variants are built upon it. The extension to LSTM and GRU will be similar. The formalization of vanila RNN is as follows:

\begin{align}
\small
\label{eq:3}
\mathbf{h}_{t+1} & = \tanh(W_h\mathbf{h}_t+W_x\mathbf{x}_{t+1}),\\
\label{eq:4}
\mathbf{y}_{t} &= f(W_o\mathbf{h}_t),
\end{align}
where $W_h\in \mathbb{R}^{m\times m}$, $W_x\in \mathbb{R}^{m\times n}$, and $W_o\in \mathbb{R}^{l\times m}$ are the coefficient matrix. The activation function $f$ is  chosen according to the task. 

As the length of the historical records varies, the RNN that learns the  vector addition should  output the cumulative sum at each step. Thus, $\mathbf{y}_t$ should be the predicted sum of input $\mathbf{x}_1, ..., \mathbf{x}_t$. The corresponding ground truth is $\sum\limits_{i=1}^t \mathbf{x}_i$.
As our goal is to learn a linear operation addition, the nonlinear layer is not necessary. Thus, we remove all the nonlinear layers and rewrite Formula~\ref{eq:3} and~\ref{eq:4} as follows:

\begin{align}
\small
\label{eq:5}
\mathbf{h}_{t+1} & = W_h\mathbf{h}_t+W_x\mathbf{x}_{t+1},\\
\label{eq:6}
\mathbf{y}_{t} &= W_o\mathbf{h}_t,
\end{align}
If we recursively apply Equations~\ref{eq:5} and \ref{eq:6}, we obtain  

\begin{align}
\small
\label{eq:7}
\mathbf{y}_{t} &=  \sum\limits_{i=1}^t W_oW_h^{t-i}W_x\mathbf{x}_i,
\end{align}
where $\mathbf{h}_0$ is the initial state which is a zero vector. Thus, the closed-form solution is $W_oW_x=\mathbb{I}^{n\times n}=\mathbb{I}^{l\times l}$ and $W_h=\mathbb{I}^{m\times m}$. This closed-form solution indicates that the vanilla RNN can  represent the vector addition without too many parameters if we can meet the constraints in the closed-form solution. A single layer RNN with  hidden state of small dimensionality has  enough representation ability. As the  nonlinear activation function may affect the learning process, we also  re-conduct the  experiments to learn the vector addition with the simplified RNN version described by Equation~\ref{eq:5} and ~\ref{eq:6}. Other configurations are the same as before. The training process is shown in  Figure~\ref{fig:no_nonlinear}. The training error  still converges around 4.   Thus, our results imply that the vanilla RNN has the ability to learn vector addition in theory,  but in practice the optimizers  cannot find this global minimal (or unable to do so in feasible time). To this end, we demonstrate the RNN based methods fail to capture PIF.

\subsubsection{How to Overcome the Difficulty}
The closed-form solution provides a direct solution to overcome this difficulty by initializing the parameters in the RNN with the optimal solution. However, the closed-form solution forces the  weighted matrices to be  correlated to each others,  which violates the effect of random and orthogonal weight initializations.  Recent literature shows that the neural networks trained by stochastic gradient descent (SGD) from random  initialization almost \textit{never suffer from} non-smoothness or non-convexity, and can avoid local  minima~\cite{goodfellow2014qualitatively}. And  orthogonal weight initializations improve the  convergence~\cite{hu2020provable}. Even though our closed-form solution is easy for the RNNs to learn  the vector addition, it brings  difficulty in training the RNNs to learn other objectives, e.g.  temporal dynamics which is also important in NBR.  We believe the solution to overcome the difficulty in training RNNs from the  optimization perspective is not trivial. So we think  PIF should be  carefully captured as it is hard to learn PIF with RNNs.  

\section{Proposed Method}
\label{sec:method}

Model-based methods have  been considered as better solution for traditional collaborative filter recommendation problem than neighbor-based methods as model-based methods can better  generalize to unseen patterns~\cite{ning2015comprehensive}. However, our empirical  results in section  ~\ref{sec:analysis_freq} show that the unseen patterns only account for a small part in the target basket. Thus, we propose to resort to the classical and direct neighbor-based methods. To  our best knowledge, even  though kNN methods have  been developed in  collaborative filter for top  $n$ recommendation~\cite{konstan1997grouplens}\cite{deshpande2004item}
 and  sequential/session  recommendation~\cite{jannach2017recurrent}, the kNN  based method for NBR has not been explored. We will leave the model-based methods as our future work.

In the following,  we introduce a simple and effective kNN based  method called  \textbf{t}emporal-\textbf{i}tem-\textbf{f}requency-based  \textbf{u}ser-KNN (TIFU-KNN).  The proposed method directly utilizes  the two important  patterns associated with PIF.  In addition, temporal dynamics is also considered to help select the items.

\begin{figure}[!t]
\centerline{\includegraphics[width=0.27\textwidth,height=0.15\textheight]{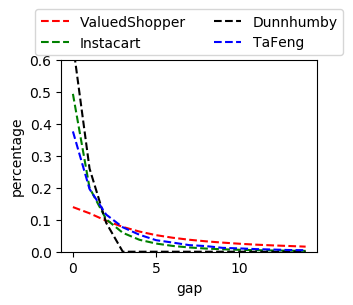}}
\caption{Gap distribution of repeated purchase.}
\label{fig:gap_distribution}
\end{figure}

\subsection{Integrating Temporal Dynamics}

% \begin{table}[ht]
% \small
% \centering
% \caption{The $s$ is  the predicted next basket size.  The $ratio$ is the ratio between the number of users whom we have the  difficulty in selecting items for and the total number of users. The $\#confused$ is the average number of items that are  difficult to be differentiated due to the same item frequency. The results indicate that the difficulty exists in a group of users and there are many options if the difficulty is encountered.}
% \begin{tabularx}{\linewidth}{XXXX}
% \toprule
% Data&$s$ &ratio &\#confused\\
% \midrule
% \multirow{2}{*}{\emph{ValuedShopper}}&10  &0.3375 &4.98\\
%     &20 &0.2696 &11.14\\
% \multirow{2}{*}{\emph{Instacart}}&10  &0.5904 &11.98\\
%     &20 &0.4562 &19.38\\
% \multirow{2}{*}{\emph{Dunnhumby}} &10  &0.5278&20.20 \\
%     &20   &0.3148&29.82 \\
% \multirow{2}{*}{\emph{TaFeng}} &10  &0.3268&22.76 \\
%     &20   &0.2037&32.23\\    
% \bottomrule
% \end{tabularx}
% \label{tab:equal_freq}
% \end{table}

\begin{figure*}[!t]
\centerline{\includegraphics[width=1\textwidth,height=0.19\textheight]{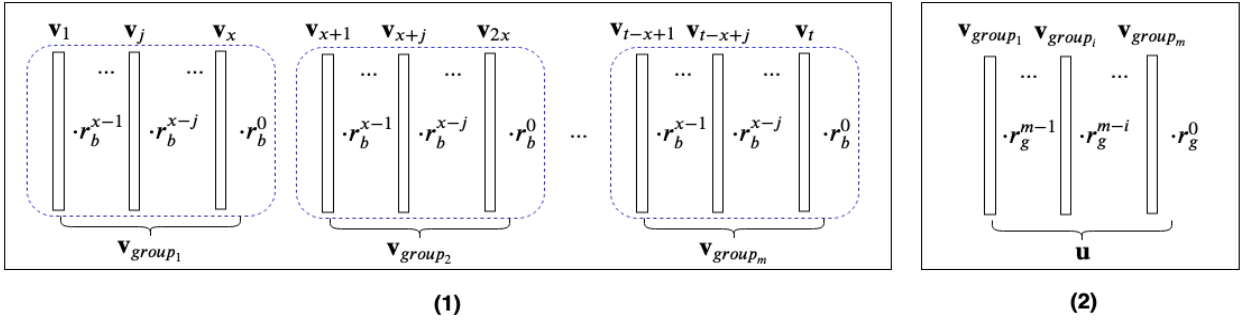}}
\caption{User vector representation generation process.}
\label{fig:user_vector}
\end{figure*}

PIF contains important information as we discuss in Section~\ref{sec:analysis_freq}. However, there is a limitation: it  cannot provide  discriminative information for items with the same item frequency. Consequently,  it is hard to  distinguish items only with item frequency, which affects both neighbors searching and items selection in kNN-based method. To address this issue, we propose to consider the temporal dynamics of the repeated purchase.  Figure~\ref{fig:gap_distribution} shows the gap distribution of repeated purchase on four data sets used in our experiments. The gap value means that after  how many baskets, the next purchase for the same item occurs again. We can observe that the short gaps dominate the repeated purchase. However, the gap distribution varies dramatically across different data sets. For ValuedShopper data set, the percentage of different gaps decreases slowly while other three data sets decreases faster. Generally, we can observe that recent purchases have more impact to trigger a repeated purchase than the behavior long time ago. Thus, we propose to assign decayed weights to the same item appearing at different time steps. The earlier the item appears, the smaller weight the item contributes to the final  frequency. We will describe the details in the next section. 

\subsection{Nearest Neighbors based Method}
\label{sec:representation}
Our kNN method consists of  two parts: the similarity calculation (between the target user and other users) and the prediction (based on the target user and his neighbors).\\ 
\textbf{User Similarity Calculation:} Considering each  user's historical records are  sequential sets of variable length, we propose to aggregate the historical records  into one vector which is easy for similarity calculation. The direct way to aggregate the  historical records is to sum them up. But this way has a limitation that we have shown in the last section. Also, it ignores that the users' preferences for products are drifting over time~\cite{koren2009collaborative}. This drifting  suggests that recent  records have more impact  than the records long time ago. Thus, we make the  items bought recently  contribute more in  the similarity calculation than the items bought long time ago. However, a single time decayed weight is not flexible to model another  property of temporal  dynamics that consecutive  steps have small changes  while  steps far from each others have large changes. To capture both temporal  dynamics, we propose to use  hierarchical time decayed weights. Our user vector representation generation process is as follows (which is shown in the Figure~\ref{fig:user_vector}): 
\\(1) We partition the historical $t$ baskets (vectors)  into $m$ groups equally. Denote the  group size as  $x=\frac{t}{m}$. The  $j$-th vector (in temporal order) within each group is  multiplied by a time-decayed weight  $r_{b}^{x-j}$, where $r_{b}$ is the time-decayed ratio within group. Then, we calculate the average vector of the weighted vectors within each group as the  corresponding group vector  $\mathbf{v}_{group}$. If the vectors cannot be equally partitioned,  the group size $x$ is calculated by $\lceil \frac{t}{m}  \rceil$ except for the first group whose size is $t-x\cdot(m-1)$. 
\\
(2) The $i$-th group vector  $\mathbf{v}_{group_i}$ is multiplied by a time-decayed weight $r_{g}^{m-i}$, where  $r_{g}$ is the  time-decayed ratio across the groups. Then, we calculate the average vector of the weighted group vectors   as the user vector representation  $\mathbf{u}$. 

After we obtain the user vector representation, we can use different methods~\cite{sarwar2001item} to calculate the similarity. Here we use the  Euclidean distance to help calculate the similarity between users. The small(large) distance means  large(small) similarity. We will leave the  exploration of other similarity functions as our future work.  We search the $k$  nearest neighbors for each target user. 
\\
\textbf{Prediction:} 
Our prediction is a  combination of  following two parts:
\begin{itemize} 
\item{Repeated purchase  component:} Denote the user representation of the target user as  $\mathbf{u_t}$. It  is corresponding to  repeated purchase  pattern. 
\item{Collaborative purchase component:}  Denote the set of  target user's nearest neighbors vector representations as  $U_{neighbor}$. Denote  the average vector of  all vectors belong to  $U_{neighbor}$ as $\mathbf{u_n}$. $\mathbf{u_n}$ is corresponding to collaborative purchase pattern. 
\end{itemize}
The final prediction is: $$\mathbf{P} = \alpha\cdot\mathbf{u_t}+(1-\alpha)\cdot\mathbf{u_n},$$ where the $\alpha$ is a hyper-parameter to balance two parts. The $s$ items  corresponding to the largest  $s$ entries in  $\mathbf{P}$ are recommended.   

% \subsection{Relation to the Existing Methods}

% \subsubsection{TIFU-KNN and userKNN}

\section{Related Work}
\label{sec:related_work}
The related works include (1) Traditional collaborative recommendation 
 methods that model user preferences without considering the 
temporal dynamics and have a set of items as the historical record; 
 (2) Sequential  recommendation methods that deal with a sequence of items or actions (each  element is an item or action) as the user profile; and (3) NBR methods that deal with a sequence of baskets (each element is a set of items) as the user profile. \\
\textbf{Traditional collaborative recommendation:}  Collaborative Filtering (CF)~\cite{ricci2011introduction} is the classical recommendation method. CF usually learns from user-item ratings matrix and predict only based on this matrix. Existing CF  methods can be classified into two categories: neighborhood- and model-based methods. Neighborhood-based methods are widely studied in traditional collaborative recommendation~\cite{ning2015comprehensive}. The  neighborhood-based methods  contain two ways:  user-based or item-based
recommendation. User-based method like GroupLens~\cite{konstan1997grouplens} predicts the  interest of a target user for an item using the ratings for this item by the most similar users. The item-based method like itemKNN~\cite{deshpande2004item} predicts the user-item  rating based on the ratings of the target user for similar
items. 
Model-based approaches use these ratings to learn a predictive
model~\cite{koren2008factorization}\cite{bell2007modeling}. Recently,  neural networks-based methods are proposed to enhance the CF  methods~\cite{he2017neural}\cite{liang2018variational}\cite{he2020lightgcn} as more nonlinear relations can be captured by neural networks.\\
\textbf{Sequential  recommendation:} The goal of sequential recommendation is to recommend the next item or action based on the past sequence of items~\cite{he2016fusing}\cite{kang2018self}\cite{he2018translation}\cite{yuan2020future}. Due to the  sequence structure in the historical records, natural language processing methods, like RNNs, attention  mechanism, and Markov chain, can be applied to model the  data~\cite{he2016fusing}\cite{kang2018self}\cite{hidasi2015session}. Session-based Recommendation also belongs to this type as each session is a short  sequence of behaviors or items~\cite{ludewig2018evaluation}\cite{jannach2017recurrent}. A kNN-based method shows competitive performance when it is compared to RNN-based method GRU4rec~\cite{jannach2017recurrent}. Our kNN-based method is different from this method in both similarity calculation step and prediction step. Also, their method cannot be directly applied to  NBR as discussed in the introduction.\\
\textbf{Next-basket recommendation:} NBR aims at predicting a set of items based on a sequence of past baskets (sets)~\cite{rendle2010factorizing}\cite{wang2015learning}\cite{yu2016dynamic}\cite{ying2018sequential}\cite{bai2018attribute}\cite{hu2019sets2sets}. The summary can be found in section~\ref{sec:existing_NB}. Unlike traditional collaborative recommendation and sequential recommendation, the study towards kNN-based method on NBR is lacked. There is no clue if this type of methods can provide better performance. Our proposed  kNN-based method fills this gap.
\\
\textbf{Difficulty in  Training RNNs:}
The vanishing and the exploding gradient problems are the well-known difficulty in training RNNs ~\cite{pascanu2013difficulty}. We present another phenomenon that it is difficult  for RNNs to learn a simple operation---vector addition. Even though we know training a deep neural network is  np-complete in the worst case~\cite{blum1989training}, the  phenomenon discovered  in this paper is different as we provide a  closed-form solution. There is a need for more  theoretical  analysis to  understanding this kind of  difficulty in training RNNs. 

\section{Experiments}
\label{sec:experiment}

In this section, we conduct extensive experiments to answer the following research questions: 
\\
\textbf{RQ1:} How is the effectiveness of the proposed methods? Can they   outperform the  state-of-the-art NBR  methods? 
\\
\textbf{RQ2:} How is the effectiveness of the temporal dynamics?
\\
\textbf{RQ3:} How is the effectiveness of each component to predict the target basket in the TIFU-KNN?
\\
\textbf{RQ4:} How do the hyper parameters affect the performance?  Does each factor in the TIFU-KNN bring benefits?

\subsection{Experimental Settings}
\subsubsection{Data sets}
\label{sec:exp_data}
We use four public data sets: 
 Dunnhumby\footnote{https://www.dunnhumby.com/careers/engineering/sourcefiles}, ValuedShopper\footnote{https://www.kaggle.com/c/acquire-valued-shoppers-challenge/overview},  Instacart\footnote{https://www.kaggle.com/c/instacart-market-basket-analysis}, and TaFeng\footnote{https://www.kaggle.com/chiranjivdas09/ta-feng-grocery-dataset}. These data sets contain the  transactions about which items are bought by  which customer at which  time. All the items bought in the same order are treated as a basket.  We remove all the customers who have baskets  less than 3 to ensure that temporal  patterns exist in the past records. In Dunnhumby, we use the 50k users sampled data. In ValuedShopper data set, we use the sampled transactions data. In Instacart and TaFeng data sets, we remove the least frequent items. The left items retain more than 95\%  item purchase of all the transactions. The  statistics of the data sets after pre-processing is shown in the Table~\ref{tab:statistic}.

\begin{table}[ht]
\small
\centering
\caption{Statistic information after pre-processing.}
\begin{tabularx}{\linewidth}{lXXXXX}
\toprule
Data &\#items &\#users &average basket size &average \#baskets /user\\
\midrule

\emph{ValuedShopper} &7,907  &10,000& 8.71& 56.85\\
\emph{Instacart} &8,000  &19,935 &8.97&7.97 \\
\emph{Dunnhumby} &4,997  &36,241 &7.33&7.99 \\
\emph{TaFeng}&12,062&13,949&6.27&5.69\\
\bottomrule
\end{tabularx}
\label{tab:statistic}
\end{table}

\subsubsection{Evaluation Protocol}

We use \textbf{recall} and \textbf{NDCG} to  evaluate our methods.
Recall is a wildly-used measurement in the  NBR~\cite{ying2018sequential}.
NDCG is a ranking based measure which takes into account the order of elements in a list~\cite{he2015trirank}. We calculate the NDCG for each basket  based on the top $s$ sorted  elements list. All the measurements are calculated across all predicted next set of items. 

% \subsubsection{Evaluation Method} 
We use the  past baskets of a given customer to  predict his/her last basket. All the data sets are partitioned across users.  The data is randomly  partitioned into 5 folds across users. And 4 folds is used for training and 1 fold is used for test. We reserve the data of 10\% users in the training set as the validation set for hyper parameters searching in all the methods.

\subsubsection{Compared  Methods}
\label{sec:baseline}
\ \\
\textbf{Simple baselines}:
\begin{itemize}

\item{Top-$n$ frequent (TopFreq):} It uses the most frequent $s$ items  that appear in all the baskets of  the training data as the predicted  next baskets for  all persons.
\item{Personalized Top-$n$ frequent (PersonTopFreq):} It uses the most frequent $s$ items that appear in the past baskets  of a given person as the prediction for the next basket. It directly use the PIF.
\end{itemize}
% \item{itemKNN:} A traditional  collaborative filter method based on kNN and item-item  similarities~\cite{sarwar2001item}. All the  items in the  historical baskets of a user are treated as the set of items that the user purchased.  We recommend the top $s$ items as the next basket. 
\textbf{Tweaked methods}:
\begin{itemize}
\item{userKNN:} It is classical collaborative filter  based on  kNN~\cite{konstan1997grouplens}. In order to apply this method, we merge  all the  items in the  historical baskets as a set of items. We recommend the top $s$ items as the next basket. This baseline can show the difference  between the  proposed method and the existing user-based kNN method.
\item{RepeatNet:} The latest  RNN-based model  for session-based recommendation which captures the repeated behaviors~\cite{ren2019repeatnet}. To apply this method to solve our problem, we transfer each basket into a sequence based on the ID's order. Then, we concatenate the sequences from  different baskets in temporal order and get a  sequence of items for each user.
\end{itemize}
\textbf{Existing NBR methods}:
\begin{itemize}

\item{FPMC:} The classical  factorization based method for next basket recommendation. It use Markov chain and factorization method to represent the past baskets~\cite{rendle2010factorizing}. Both sequential behaviors and users' personal tastes are taken into account for prediction. 
\item{DREAM:} A  deep model based on embedding and RNN for next basket  recommendation~\cite{yu2016dynamic}. It considers  personal dynamic interests at different time and the global interactions of all baskets of the user over time. 
\item{SHAN:} A deep model based on  hierarchical attention networks~\cite{ying2018sequential} . It partitions  the historical baskets into long-term and short-term parts to learn the long-term preference and short-term preference  based on the corresponding items attentively. It can be directly applied in  NBR and sequential/session-based recommendation as it treats the historical records as two sets. 
\item{Sets2Sets:} The state-of-the-art  end-to-end method for  following multiple baskets prediction based on RNN~\cite{hu2019sets2sets}. Repeated purchase pattern is also integrated into the method. 
\end{itemize}

We focus on comparing with existing
 NBR methods. Other  techniques used in   top $n$  recommendation and  sequential/session-based recommendation are not the focus of this paper.

\begin{table}[ht]
\small
\centering
\caption{Parameters of our methods in different data sets.}
\begin{tabularx}{\linewidth}{lXXXXXX}
\toprule
\textbf{Data}&$k$ &$m$  &$r_b$ &$r_g$  &$\alpha$\\
\midrule
\emph{ValuedShopper} &300 &7  &1&0.6 &0.7 \\
\emph{Instacart}&900 & 3 &0.9  &0.7  & 0.9\\
\emph{Dunnhumby} &900 &3 &0.9  &0.6 &0.2 \\
\emph{TaFeng} &300 &7 &0.9  &0.7 &0.7 \\
\bottomrule
\end{tabularx}
\label{tab:param}
\end{table}

\begin{table*}[ht]

\small
%\begin{adjustwidth}{-1cm}{}
\caption{Comparison with different methods. The \textbf{bold} is the maximum in (a)-(b). The \underline{underline} is the maximum in (c)-(h).}
\centering
\resizebox{1\textwidth}{!}{
\begin{tabular}{lccccccccc|ccc}
%\begin{tabular}{llllllllllllllll}
\toprule
 \multirow{2}{*}{Data}&\multirow{2}{*}{Metric} &(a)&(b)&(c)&(d)&(e)&(f)&(g)&(h)&(i)&\multicolumn{2}{c}{improvement vs.}\\
  &&TopFreq&PersonTopFreq&userKNN&RepeatNet&FPMC&DREAM&SHAN&Sets2Sets&TIFU-KNN&(a)-(b)&(c)-(h)\\

\midrule
\multirow{4}{*}{\emph{ValuedShopper}}&recall@10 &0.0982&\textbf{0.2109}&0.0988&0.1031&0.0951&0.0991&0.0847&\underline{0.1259}&0.2162&2.5\%&71.7\%\\
&recall@20 &0.0904&\textbf{0.2969}&0.1329&0.1485&0.1391&0.1448&0.1220&\underline{0.1774}&0.3028&2.7\%&70.6\%\\
&NDCG@10 &0.0779&\textbf{0.2128}&0.1415&0.1439&0.1188&0.1231&0.1032&\underline{0.1626}&0.2171&2.1\%&33.5\%\\
&NDCG@20  &0.0904&\textbf{0.2544}&0.1662&0.1693&0.1253&0.1287&0.1074&\underline{0.1884}&0.2589&1.7\%&37.4\%\\
\midrule
\multirow{4}{*}{\emph{Instacart}}&recall@10 &0.0724&\textbf{0.3426}&0.0720&0.2107&0.0763&0.0866&0.0902&\underline{0.3021}&0.3952&15.3\%&30.8\%\\
&recall@20 &0.1025&\textbf{0.4652}&0.1260&0.2637&0.1073&0.1128&0.1246&\underline{0.3654}&0.4875&4.8\%&33.4\%\\
&NDCG@10 &0.0641&\textbf{0.3618}&0.1020&0.2285&0.0946&0.1063&0.1152&\underline{0.3487}&0.3825&5.7\%&9.6\%\\
&NDCG@20  &0.0689&\textbf{0.4155}&0.1394&0.2513&0.0992&0.1157&0.1212&\underline{0.3626}&0.4384&5.5\%&20.9\%\\
\midrule
\multirow{4}{*}{\emph{Dunnhumby}}&recall@10 &0.0819&\textbf{0.1853}&0.1135&0.1324&0.0919&0.0915&0.1007&\underline{0.2068}&0.2087&12.6\%&0.9\%\\
&recall@20 &0.1077&\textbf{0.2366}&0.1648&0.1989&0.1186&0.1087&0.1201&\underline{0.2653}&0.2692&13.7\%&1.4\%\\
&NDCG@10 &0.0601&\textbf{0.1771}&0.1707&0.1545&0.1025&0.1009&0.1149&\underline{0.2134}&0.1983&11.9\%&-7.0\%\\
&NDCG@20  &0.0609&\textbf{0.2016}&0.2052&0.1732&0.1057&0.1022&0.1167&\underline{0.2385}&0.2302&14.1\%&-3.5\%\\
\midrule
\multirow{4}{*}{\emph{TaFeng}}&recall@10 &\textbf{0.0773}&0.0704&0.1089&0.0645&0.0868&0.0902&0.0878&\underline{0.1190}&0.1301&33.7\%&9.3\%\\
&recall@20 &0.1151&\textbf{0.1203}&0.1278&0.0919&0.1056&0.1149&0.1065&\underline{0.1767}&0.1810&50.4\%&2.4\%\\
&NDCG@10 &0.0519&\textbf{0.0766}&0.0832&0.0592&0.0667&0.0763&0.0813&\underline{0.0844}&0.1011&31.9\%&8.4\%\\
&NDCG@20  &0.0608&\textbf{0.0896}&0.1064&0.0679&0.0743&0.0841&0.0892&\underline{0.1071}&0.1206&34.5\%&12.6\%\\
\bottomrule
\end{tabular}
}
\label{tab:comp1}
%\end{adjustwidth}
\end{table*}

We tune the hyper parameters in all the compared methods with grid search to achieve their best performance. For userKNN, the number of nearest neighbors is searched from the set of values [100, 300,500, 700, 900, 1100, 1300]. For FPMC, the dimension of factor is searched from the set of values [16, 32, 64, 128]. For RepeatNet, DREAM, SHAN, and Sets2Sets, the  embedding size is searched from the set of values [16, 32, 64, 128].

\subsubsection{Configuration of the Proposed  Method}
We  perform an extensive search over the parameter space to achieve the best performance on the validation  set. The number of nearest neighbors $k$ is chosen from the set of values [100, 300, 500, 700, 900, 1100, 1300]. The the within-basket time-decayed ratio  $r_b$ and the group  time-decayed ratio $r_g$ are chosen from the set of values [0.1, 0.2, 0.3, 0.4,  0.5, 0.6, 0.7, 0.8, 0.9, 1].  The fusion weight  $\alpha$ is searched from the set of values [0, 0.1, 0.2, 0.3, 0.4, 0.5, 0.6, 0.7, 0.8, 0.9, 1]. The  number of groups  $m$ is  searched from the set of values [3, 7, 11, 15, 19, 23] in ValuedShopper data set and from the set of values [2, 3, 4, 5, 6, 7] in other data sets,  respectively. The parameters associated with the  results  reported in  the methods  comparison are shown in the Table~\ref{tab:param}.

\subsection{Performance Comparison (RQ1)}

The comparisons with baselines and existing  methods are shown in the Table~\ref{tab:comp1}. Several observations can be  made. First, the simple  top-$n$ frequent  baseline achieves  reasonable  performance  compared to other  existing methods in recall.  It indicates that  some popular items are commonly purchased by  different users. But this simple baseline almost produces the worst performance across different data sets. It implies that users also have their distinct items which cannot be obtained through this simple baseline. 

Second, personalized top-$n$ frequent  method achieves competitive performance across all data sets. This verifies that the impact of the repeated purchase pattern from  the target users plays an  important role in the  prediction. 

Third,  the existing NBR methods (excluding Sets2Sets) is  surpassed by the baseline  personalized top-$n$  frequent  method in ValuedShopper,  Instacart, and Dunnhumby data sets  by a large margin. The  reason is that existing methods  (excluding  Sets2Sets) cannot capture  PIF. Even though Sets2Sets  captures  PIF explicitly, it is still worse than the personalized top-$n$  frequent method in first two data sets.  We believe the reason is that the  learned coefficients cannot perfectly control Sets2Sets to rely on the PIF and RNN based module. 

Fourth, the tweaked top-$n$ recommendation method userKNN and session-based recommnedation method RepeatNet are worse than  the state-of-the-art  method Sets2Sets. The  reason is that they ignore the important information existing in the sequential sets. For userKNN, it discards the PIF. RepeatNet outperforms other method without repeated purchase  pattern when the repeated purchase affects a lot in the data (excluding TaFeng data set). But it performs worse than other methods with repeated purchase pattern (personalized  top-$n$ frequent baseline and Sets2Sets) as it also captures the non-existing order among the items within each basket. 

Fifth, the proposed  TIFU-KNN is better than other methods in  ValuedShopper,  Instacart and TaFeng data sets,  which verifies the  superiority of the proposed method. There is an exception in Dunnhumby data set that Sets2Sets  achieves better NDCG  while the proposed method achieves a little better recall. We believe the reason is that the embedding method used in  Sets2Sets can help generalize to some unseen user-item patterns in the training set beyond the repeated purchase pattern and collaborative purchase pattern. It is consistent with our analysis in the section~\ref{sec:analysis_freq} that Dunnhumby has much more unseen patterns than other three data sets.

\begin{figure}[!t]
\centerline{\includegraphics[width=0.5\textwidth,height=0.3\textheight]{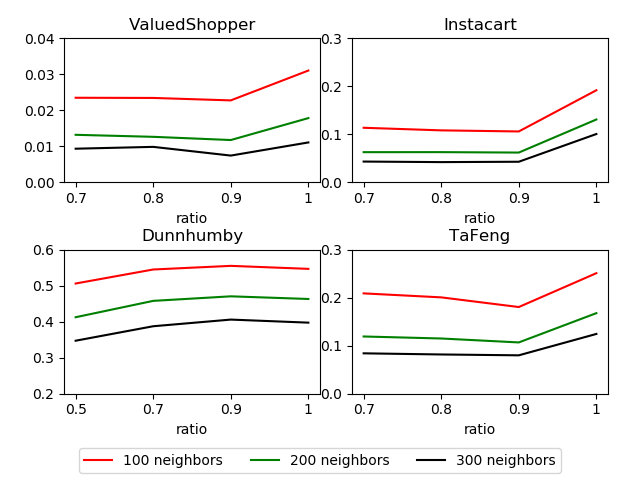}}
\caption{$Recall_{\overline{P}\cap \overline{N}}$ distribution on different data sets. The $r_b$ and $r_g$ are set to the same ratio. Different lines represent different numbers of neighbors.}
\label{fig:neg_recall}
\end{figure}

\subsection{Effectiveness of Temporal Dynamics (RQ2)}

In this section, we  investigate if the temporal  dynamics brings positive effectiveness. For  simplicity, we set the time decayed weights $r_b$ and $r_g$ to the same ratio. We  denote the set of the nonzero entries in repeated purchase component and collaborative purchase component as $P$ and $N$, respectively. Then, we use  the items in $\overline{P} \cap \overline{N}$ to retrieve items in the target basket and calculate the recall $Recall_{\overline{P}\cap \overline{N}}$. $Recall_{\overline{P}\cap \overline{N}}$ quantifies  the amount of unseen patterns. The small value  means large coverage by the repeated purchase pattern and collaborative purchase pattern. From 
Figure~\ref{fig:neg_recall}, we can observe that without the temporal dynamics, which is represented by $ratio=1$,  the proposed method usually has the largest number of unseen patterns. It implies that the temporal dynamics can help reduce the unseen patterns as better neighbors are searched.

% \subsection{Generalization vs our method}

\subsection{Effectiveness of Different Components (RQ3)}
In this section, we investigate the  contributions from two patterns associated with PIF.  The  comparison between  our full TIFU-KNN and a single component as prediction is shown in the Table~\ref{exp:comparison3}. Obviously, the combination achieves the best performance, which verifies the effectiveness of the  combination of  the target user's  PIF and  the most similar users' PIF. Also, we  observe that target user's repeated purchase  pattern dominates the  prediction. The  collaborative  purchase pattern provides   discriminative information to further improve the prediction.

\begin{table}[ht]
\small
\caption{The effect of each component in the TIFU-KNN.}
\resizebox{0.48\textwidth}{!}{
\begin{tabular}{l|ccc|ccc}
\toprule 
&&recall@10& &&NDCG@10&\\
 Data &$\small \mathbf{u_{t}}$& $\small \mathbf{u_{n}}$& $\small \mathbf{u_{t}\&u_{n}}$&$\small \mathbf{u_{t}}$& $\small \mathbf{u_{n}}$& $\small \mathbf{u_{t}\&u_{n}}$\\ 
 \hline \emph{ValuedShopper}  &0.1801	&0.1251	&\textbf{0.2161}

 &0.1716&0.1287	&\textbf{0.2171}

\\
   \emph{Instacart} &0.3698	&0.1290	&\textbf{0.3952}	&0.3686	&0.1381	&\textbf{0.3825}	
\\
   \emph{Dunnhumby} &0.2070	&0.1344	&\textbf{0.2087}		&0.1968	&0.1270	&\textbf{0.1983}

\\
   \emph{TaFeng} &0.0921	&0.0904	&\textbf{0.1301}		&0.0891	&0.0766	&\textbf{0.1011}

\\
% \hline
% &&recall@20& &&NDCG@20&\\
%  Data &$\small \mathbf{P_{t}}$& $\small \mathbf{P_{n}}$& $\small \mathbf{P_{t}\&P_{n}}$&$\small \mathbf{P_{t}}$& $\small \mathbf{P_{n}}$& $\small \mathbf{P_{t}\&P_{n}}$\\ 
%  \hline ValuedShopper &0.2591	&0.1745	&\textbf{0.3033}

%  &0.2095&0.1520	&\textbf{0.2599}

% \\
%   Instacart &0.4721	&0.1795	&\textbf{0.4841}	&0.4234	&0.1631	&\textbf{0.4384}	
% \\
%   Dunnhumby &0.2689	&0.1706	&\textbf{0.2692}		&0.2280	&0.1424	&\textbf{0.2302}

% \\
\bottomrule
\end{tabular}
}
\label{exp:comparison3}
%\end{adjustwidth}
\end{table}

\subsection{Sensitivity of the Hyperparameters (RQ4)}

\begin{figure}[!t]
\centerline{\includegraphics[width=0.5\textwidth,height=0.14\textheight]{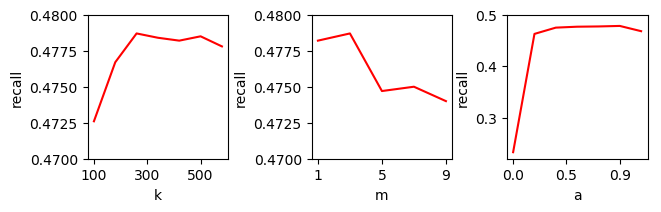}}
\caption{Sensitivity of hyperparameters: the number of nearest neighbors $k$, the number of groups $m$, and the combining weight  $\alpha$ at Instacart data set.}
\label{fig:parameter}
\end{figure}

In this section, we investigate how the hyper parameters affect the performance.   When we investigate on one or two parameters,  we set other parameters just as the value shown in Table~\ref{tab:param}. We report the recall on the test set in Instacart data set when the  predicted basket size $s=20$. The results are shown in the Figure~\ref{fig:parameter}. As  the decayed weights $r_b$ and $r_g$ may have  some correlation, we investigate  them together and the results are shown in the Table~\ref{tab:param_sensitive}. We have several observations. 
First, all the hyperparameters should be chosen with a proper value in order to achieve the best performance.  Second, the  parameters  selected with the validation set are close to the optimal  configuration for the test set. Third, the two time decayed weights $r_b$ and $r_g$ should both be smaller than 1 in order to achieve the best performance. It verifies that our two-level decayed weight design is better than any single  decayed weight.

\begin{table}[ht]
\small
%\begin{adjustwidth}{-1cm}{}
\caption{Sensitivity of hyperparameters: time-decayed ratio $r_b$ within each group and time-decayed ratio  $r_g$ across the groups at Instacart data set.}
\centering
\begin{tabular}{ccccccc}
\toprule 

recall&&&$\mathbf{r_b}$&&&\\
 $\mathbf{r_g}$&0.1&0.3&0.5&0.7&0.9&1\\
 \hline
 0.1&0.4504
    &0.4491
    &0.4474
    &0.4487	&0.4497
    &0.4505

\\
 0.3&0.4754
	&0.4755
	&0.4751
	&0.4744
	&0.4694
    &0.4692

\\
 0.5&0.4786
	&0.4783
	&0.4783
	&0.4785
	&0.4782
    &0.4759

\\
 0.7&0.4837
	&0.4834
	&0.4832
	&0.4831
	&0.4841
    &0.4825

\\
 0.9&0.4869
	&0.4872
	&0.4874
    &0.4873
	&0.4878
    &0.4872

\\
 1&0.4780
	&0.4784
	&0.4778
	&0.4775
	&0.4772
    &0.4788

\\
\bottomrule
\end{tabular}
\label{tab:param_sensitive}
%\end{adjustwidth}
\end{table}

\section{Conclusion}
\label{sec:conclusion}

In this paper, we introduce a simple kNN-based method\footnote{The code is available at~\url{https://github.com/HaojiHu/TIFUKNN}.}. Despite its simplicity, the proposed method generally outperforms the state-of-the art deep learning based methods.  We study the reason why RNNs cannot  approximate vector addition well, which provides the  insight why our proposed method can outperform existing methods. Even though the deep learning  model has strong representation  power, there is no guarantee that we can find the solution which meets our expectation due to the complexity of non-convex  optimization in RNNs. We believe this difficulty is  different from the  well-known  vanishing and the exploding gradient  problems  ~\cite{pascanu2013difficulty}.  More theoretical analysis is needed. A new optimizer that  has the theoretical  guarantee to find the global optimal like~\cite{wang2019learning} is also  needed for RNNs. 

Beyond this work, we believe that there are two  directions that deserve to be explored. First, a direct  extension is whether there are other commonly-used functions which are hard to be learned by existing widely-used deep  models. This direction can help us better understand how to apply deep learning based  methods in recommendation systems as we observe that recent  publications~\cite{ludewig2019performance} show a worry about the  unclear progress  in  sequential/session-based recommendation.  We  believe  different types  of methods should  have  different advantages in  different tasks and data sets. And a deep understanding about the boundary of the deep learning methods can bring benefits not only to recommendation systems but also to other machine learning areas. Second, another direct extension is to investigate if there are other patterns associated with PIF or other patterns that are associated with  different types of item frequency, e.g., global item frequency, local item frequency (the item  frequency associated with a small group of users or a small group of items), and inverse item  frequency~\cite{sharma2019adaptive}. 
\\
\textbf{Acknowledgement}:
This research was supported  part by NSF under grants CNS-1814322, CNS-1831140,  CNS-1901103, US DoD DTRA DTRA  grant HDTRA1-14-1-0040, and National Natural Science Foundation of China (61972372,  U19A2079). Also, thanks to the  constructive suggestions from the  reviewers.

\bibliographystyle{ACM-Reference-Format}
\bibliography{icdm19}

\end{document}